\documentclass[12pt]{article}

\usepackage{cite} 
\usepackage{epsfig} 
\usepackage{amssymb} 
\usepackage{amsmath} 
\usepackage{latexsym} 

\renewcommand{\d}{{\mathrm{d}}}
\newcommand{\x}{{\mathrm{\bf x}}}
\newcommand{\A}{{\mathrm{\bf A}}}
\newcommand{\R}{{\mathrm{\bf R}}}
\newcommand{\re}{\mathop{\mathrm{Re}}\nolimits}

\begin{document}

\boldmath
\title{
\vskip-3cm{\baselineskip14pt
\centerline{\normalsize DESY 08-037\hfill ISSN 0418-9833}
\centerline{\normalsize arXiv:0811.1364 [hep-ph]\hfill}
\centerline{\normalsize November 2008\hfill}}
\vskip1.5cm
\bf Two-loop electroweak fermionic corrections to
$\sin^2\theta_{\mathrm{eff}}^{b\overline{b}}$} 
\unboldmath

\author{M.~Awramik$^{a,b}$, M.~Czakon$^{c,d}$, A.~Freitas$^e$,
B.A.~Kniehl$^a$\\
\\
{\normalsize\it $^a$ II. Institut f\"ur Theoretische Physik,
Universit\"at Hamburg,}\\
{\normalsize\it Luruper Chaussee 149, 22761 Hamburg, Germany}\\
\\
{\normalsize\it $^b$ Institute of Nuclear Physics PAN,}\\ 
{\normalsize\it ul.\ Radzikowskiego 152, 31342 Krak\'ow, Poland}\\
\\
{\normalsize\it $^c$ Institut f\"ur Theoretische Physik und Astrophysik,  
Universit\"at W\"urzburg,}\\
{\normalsize\it Am Hubland, 97074 W\"urzburg, Germany}\\
\\
{\normalsize\it $^d$ Institute of Nuclear Physics, NCSR ``DEMOKRITOS'',}\\
{\normalsize\it 15310 Athens, Greece}\\
\\
{\normalsize\it $^e$ Department of Physics \& Astronomy, University of
Pittsburgh,}\\
{\normalsize\it Pittsburgh, PA 15260, USA}}

\date{}

\maketitle

\begin{abstract}
  We present the first calculation of the 
  two-loop electroweak fermionic correction to the flavour-dependent 
  effective weak-mixing angle for bottom quarks, $\sin^2\theta_{\mathrm{eff}}^{b\overline{b}}$. 
  For the evaluation of the missing two-loop vertex diagrams, 
  two methods are employed, one based on a semi-numerical   
  Bernstein-Tkachov algorithm and the second on 
  asymptotic expansions in the large top-quark mass. A third method based on 
  dispersion relations is used for checking the basic loop integrals. 
  We find that for small Higgs-boson mass values, 
  $M_H \propto 100$~GeV, the correction is sizable, of order 
  ${\cal O}(10^{-4})$.

\medskip


\end{abstract}

\newpage

\section{Introduction} 
%
Experiments at LEP, SLC and Tevatron have provided  
a large number of high-precision data,  
which, being supplemented by  
detailed studies of higher-order corrections,  
allow to probe the Standard Model at the loop level and subsequently  
to predict the mass of the Higgs boson.  
In this context,  
the leptonic effective weak-mixing angle,
$\sin^2\theta_{\mathrm{eff}}^{\mathrm{lept}}$,  
plays the most crucial role. 
It can be defined through the effective vector and 
axial-vector couplings, $g_V^l$ and $g_A^l$, of the $Z$ boson to leptons ($l$) at 
the $Z$-boson pole, 
\begin{equation} 
  \sin^2\theta_{\mathrm{eff}}^{\mathrm{lept}} = \frac{1}{4} 
  \left(1 + \re\frac{g_V^l}{g_A^l}\right). 
\end{equation} 
The effective weak-mixing angle can be related to the on-shell  
Weinberg angle, $\sin^2\theta_w$, as
\begin{equation} 
  \sin^2\theta_{\mathrm{eff}}^{\mathrm{lept}} = \sin^2\theta_w \, \kappa,  
  \label{eq:sin} 
\end{equation} 
where $\sin^2\theta_w = 1 - M_W^2/M_Z^2$ and $\kappa = 1 + \Delta \kappa$. At tree 
level, $\Delta \kappa = 0$ and $\sin^2\theta_{\mathrm{eff}}^{\mathrm{lept}} = \sin^2\theta_w$. The form factor 
$\Delta \kappa$ incorporates the higher-order loop corrections. 
Usually, the $W$-boson mass, $M_W$, is not treated as an input parameter 
but it is calculated from the Fermi constant, $G_\mu$, which is precisely known  
from the muon lifetime. The relation between $M_W$ and $G_\mu$ can be cast 
in the form 
\begin{equation} 
  M_W^2 \left(1 - \frac{M_W^2}{M_Z^2}\right) = 
  \frac{\pi \alpha}{\sqrt{2}G_\mu} \left(1 + \Delta r\right),  
  \label{eq:delr} 
\end{equation} 
where the quantity $\Delta r$ \cite{Sirlin:1980nh} contains all higher-order
corrections.  
The presently most accurate calculation of the $W$-boson mass includes full two-loop  
and leading higher-order corrections \cite{mw}. 
On the other hand, the quantity $\kappa$ in Eq.~(\ref{eq:sin}) incorporates all 
corrections to  the form factors of the $Zl\overline{l}$ vertex.  Recently, the 
calculation of the two-loop electroweak corrections has been  
completed \cite{sineff,Awramik:2004qv,sw2,sinbos,sineffTot}.  The uncertainty on 
$\sin^2\theta_{\mathrm{eff}}^{\mathrm{lept}}$ due to unknown higher orders has been estimated to be 0.000047,  
which is substantially smaller than the error of the current 
experimental value $\sin^2\theta_{\mathrm{eff}}^{\mathrm{lept}} = 0.23153\pm 0.00016$ \cite{Aleph}, but still larger 
than the expected precision, $1.3\times 10^{-5}$, of a future high-luminosity linear 
collider running at the $Z$-boson pole \cite{gigaz}. 
 
The experimental value for $\sin^2\theta_{\mathrm{eff}}^{\mathrm{lept}}$ is determined from six asymmetry measurements,  
$\mathcal{A}_{FB}^{0,l}$,  
$\mathcal{A}_l(P_\tau)$,   
$\mathcal{A}_l(\mathrm{SLD})$,  
$\mathcal{A}_{FB}^{0,b}$,   
$\mathcal{A}_{FB}^{0,c}$, and 
$\mathcal{Q}_{FB}^{\mathrm{had}}$.     
Of those, the average leptonic and hadronic measurements  
differ by 3.2 standard deviations, which is one of the largest discrepancies 
within the Standard Model.  
The main impact stems from two measurements, the  
left-right asymmetry with a polarised electron beam at SLD,  
$\mathcal{A}_{LR}^{0}$, and  
the forward-backward asymmetry for bottom quarks at LEP,  
$\mathcal{A}_{FB}^{0,b}$. 
On the experimental side, the only possible source of this discrepancy  
are uncertainties in external input parameters, in particular parameters 
describing the production and 
decay of heavy-flavoured hadrons; 
see Section 5 of Ref.~\cite{Aleph} for a discussion. 
However, the interpretation of the asymmetry measurements in terms  
of $\sin^2\theta_{\mathrm{eff}}^{\mathrm{lept}}$ requires also some theoretical input.  
The leptonic asymmetries depend on lepton couplings only and 
can be translated straightforwardly into the leptonic effective weak-mixing 
angle, with small corrections due to $s$- and $t$-channel photon exchange. 
By contrast, the hadronic observables,  
$\mathcal{A}_{FB}^{0,c}$,
$\mathcal{A}_{FB}^{0,b}$ and 
$\mathcal{Q}_{FB}^{\mathrm{had}}$,  
depend on the quark couplings, $g_{V,A}^q$. 
These couplings are associated with a flavour-dependent hadronic  
effective weak-mixing angle,  $\sin^2\theta_{\mathrm{eff}}^{q\overline{q}}$, 
\begin{equation} 
  \sin^2\theta_{\mathrm{eff}}^{q\overline{q}} = \frac{1}{4|Q_q|} 
  \left(1 + \re\frac{g_V^q}{g_A^q}\right), 
  \qquad q=d,u,s,c,b. 
\end{equation} 
The forward-backward pole asymmetry of a quark $q$, $\mathcal{A}_{FB}^{0,q}$,
is related to the effective couplings, $g_V^f$ and  $g_A^f$, and the effective
weak-mixing angle, $\sin^2\theta_{\mathrm{eff}}^{q\overline{q}}$, by%
\footnote{Owing to the non-zero bottom-quark mass, the $Zb\overline{b}$ vertex
also has a
scalar part, besides the vector and axial-vector parts. We checked
explicitly that the contribution of this scalar form factor to
$\mathcal{A}_{FB}^{0,b}$ is more than a factor 1000 smaller than the current
experimental uncertainty and thus truly negligible.}
\begin{equation}  
\mathcal{A}_{FB}^{0,q} = \frac{3}{4}  \mathcal{A}_e\mathcal{A}_q, 
\end{equation}  
with 
\begin{equation} 
\mathcal{A}_f = \frac{2 g_V^f g_A^f}{(g_V^f)^2 + (g_A^f)^2} 
= \frac{1 - 4 |Q_f| \sin^2\theta_{\mathrm{eff}}^{f\overline{f}}}
{1-4 |Q_f| \sin^2\theta_{\mathrm{eff}}^{f\overline{f}} + 8 Q_f^2 
\sin^4\theta_{\mathrm{eff}}^{f\overline{f}}}. 
\end{equation} 
At tree level, $\sin^2\theta_{\mathrm{eff}}^{q\overline{q}}$ and $\sin^2\theta_{\mathrm{eff}}^{\mathrm{lept}}$ are identical, but the relations between 
these quantities receive sizable radiative corrections that need to be included 
in the analysis. 
Note that, due to the small electric charge of the bottom quark, $Q_b = -1/3$, the parameter 
$\mathcal{A}_b$ is close to 1, and $\mathcal{A}_{FB}^{0,b}$ is only weakly sensitive to $\sin^2\theta_{\mathrm{eff}}^{b\overline{b}}$. 
Therefore, it seems unlikely that the discrepancy between 
$\mathcal{A}_{LR}^0$ and $\mathcal{A}_{FB}^{0,b}$ could be explained by 
radiative corrections. Nevertheless, the theoretical prediction for $\sin^2\theta_{\mathrm{eff}}^{b\overline{b}}$ 
enters in the Standard-Model fits through several observables, so that a precise prediction 
of this quantity is important for a robust analysis. 
 
For all fermions except bottom quarks, the known radiative corrections to  
$\sin^2\theta_{\mathrm{eff}}^{f\overline{f}}$ include at least two-loop fermionic electroweak contributions and 
some leading higher-order corrections; see Ref.~\cite{sineffTot} for details.  
However, for the $Zb\overline{b}$ vertex only one-loop corrections, leading 
two-loop corrections for large values of the top-quark mass of 
$\mathcal{O}(\alpha^2 m_t^4)$, and two- and three-loop QCD corrections have been calculated \cite{Barbieri:1992nz} and included 
in the {\sc Zfitter} program 
\cite{Bardin:1992jc} (see also the new program {\sc Gfitter} \cite{Flaecher:2008zq}), which is widely used for 
global Standard-Model fits. The remaining two-loop electroweak corrections beyond the $\mathcal{O}(\alpha^2 
m_t^4)$ contributions are still unknown, although they are expected to be larger 
than the $\mathcal{O}(\alpha^2 m_t^4)$ term, based on experience from $\sin^2\theta_{\mathrm{eff}}^{\mathrm{lept}}$. 
As a result, the present treatment of higher-order electroweak corrections leads 
to inconsistencies, for example in $\mathcal{A}_{FB}^{0,b}$, since the corrections to $\sin^2\theta_{\mathrm{eff}}^{\mathrm{lept}}$ and 
$\mathcal{A}_e$ include two-loop  
and leading three-loop corrections that are absent for $\sin^2\theta_{\mathrm{eff}}^{b\overline{b}}$ and 
$\mathcal{A}_b$ (see recent discussion in Ref.~\cite{Freitas:2004mn}).  
 
In this paper, the part of the missing two-loop corrections to $\sin^2\theta_{\mathrm{eff}}^{b\overline{b}}$ with closed fermion 
loops is presented. We begin by explaining the techniques employed for the 
calculation in the next section. In Section~\ref{results}, numerical results for 
$\sin^2\theta_{\mathrm{eff}}^{b\overline{b}}$ are given before the summary in Section~\ref{concl}.

\section{Outline of the calculation} 
 
\subsection{General approach} 
 
We work in the Standard Model and adopt the on-shell renormalisation scheme,  
which relates the renormalised masses and couplings to physical observables. 
Details on the renormalisation scheme and explicit expressions for the relevant 
counterterms can be found in Refs.~\cite{sineffTot,muon}. 
For the loop integrations, we employ dimensional regularisation. The problem of  
$\gamma_5$ matrices in two-loop vertex diagrams with fermion triangle sub-loops 
is treated in the same way as in Refs.~\cite{sineffTot,sineff,Awramik:2004qv}, by evaluating the 
finite non-anticommutative contribution from $\gamma_5$ to the vertex diagrams 
in four dimensions. 
Most aspects connected with the calculation of the  
effective weak-mixing angle for the $Zb\bar{b}$ vertex are the same as  
for the leptonic effective weak-mixing angle and are discussed in detail in  
Ref.~\cite{sineffTot}. 
 
The contributions for the two-loop renormalisation terms are identical 
to the case of $\sin^2\theta_{\mathrm{eff}}^{\mathrm{lept}}$, with the exception of the two-loop bottom-quark
wave-function  
counterterm, which involves new self-energy diagrams with internal top-quark 
propagators; 
the first terms of this quantity are given in Ref.~\cite{Butenschoen:2007hz}.   
For the two-loop $Zb\overline{b}$ vertex corrections, on the other hand, 
a number of new three-point diagrams need to be computed. 
In general, electroweak two-loop corrections can be divided into two groups,  
which are separately finite and gauge invariant:  
fermionic corrections (with at least one closed fermion loop) and  
bosonic corrections (without any closed fermion loops). 
In this article, we focus on the fermionic diagrams as a first step. 
 
For the purpose of this calculation, all light-quark masses are neglected in the 
two-loop diagrams, including the bottom-quark mass. 
As a result, for many diagrams, known results from the $\sin^2\theta_{\mathrm{eff}}^{\mathrm{lept}}$ calculation 
can be used \cite{sineffTot,sineff,Awramik:2004qv}. The loop integrals for diagrams with closed 
massless-fermion loops are given in analytical form, while 
large-mass expansions were employed for diagrams with top quarks in the 
loops. 
 
\begin{figure}[tb] 
  \begin{center} 
    \raisebox{9cm}{} 
    \psfig{figure= 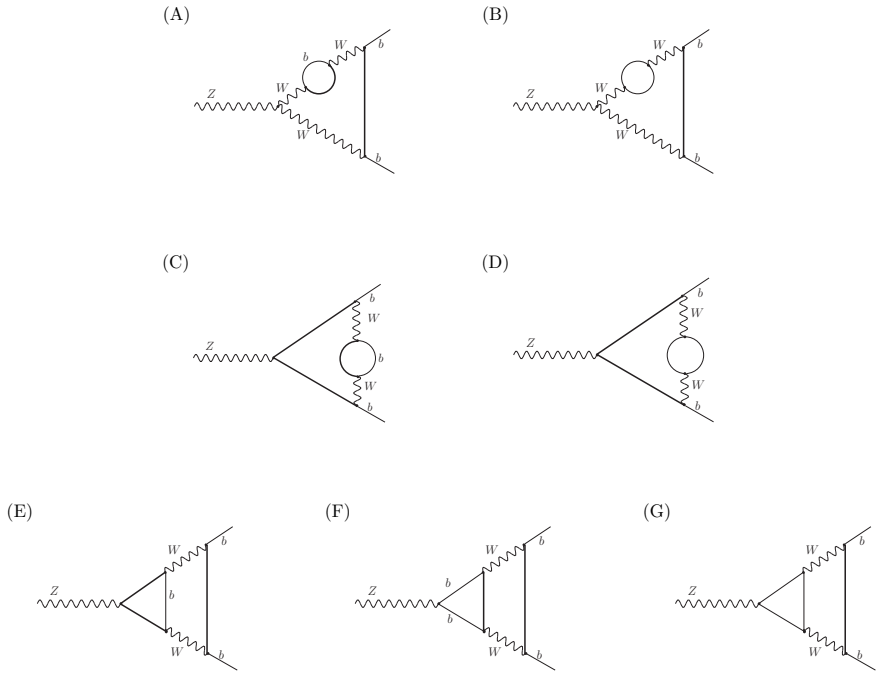, width=12cm}  
  \end{center} 
  \caption{ 
    Set of Feynman diagrams required for the calculation of the
    fermionic two-loop corrections to the $Zb\overline{b}$ vertex, but absent
    in the $\sin^2\theta_{\mathrm{eff}}^{\mathrm{lept}}$ case.  
    Thick solid lines denote top-quark propagators, while thin lines represent light 
    fermions. 
    \label{diags}} 
\end{figure}

However, the two-loop corrections to $\sin^2\theta_{\mathrm{eff}}^{b\overline{b}}$ include a new group of integrals 
that were not covered in previous calculations of $\sin^2\theta_{\mathrm{eff}}^{\mathrm{lept}}$, stemming from 
diagrams with internal $W$-boson and top-quark propagators; see Fig.~\ref{diags}. 
The computation of these diagrams will be discussed in detail in the following 
subsections. 
%
%
The two-loop diagrams are computed with several independent methods, so that 
cross checks can be performed. The first method, based on the observation that 
all new diagrams in Fig.~\ref{diags} include internal top-quark propagators, 
uses asymptotic expansions for large top-quark mass. This method was already
employed successfully for the calculation of $\sin^2\theta_{\mathrm{eff}}^{\mathrm{lept}}$ \cite{sineffTot}. 
For references on the subject, we refer the reader to Ref.~\cite{Smirnov:2004ym}. 
 
Secondly, we develop a code for the evaluation  
of Feynman diagrams with a semi-numerical method, based on the Bernstein-Tkachov (BT) method of Ref.~\cite{Tkachov}. 
This method had already been used previously for one-loop problems 
\cite{Bardin:2000cf}. 
In a recent series of papers \cite{Passarino}, it was extended to 
general two-loop vertices, and 
some applications to two-loop problems are already known: 
the leptonic effective weak-mixing angle was presented in  
Ref.~\cite{sw2} and corrections to the $H \to \gamma \gamma$ decay width
in Ref.~\cite{Passarino:2007fp}.  
 
Finally, we use another semi-numerical method based on dispersion relations 
\cite{intnum}, which was also used previously for $\sin^2\theta_{\mathrm{eff}}^{\mathrm{lept}}$ \cite{sineffTot}. 
This method allows us to evaluate all self-energy diagrams, the vertex diagrams in 
Figs.~\ref{diags}(A)--(D), as well as the scalar integrals with the topology of 
Figs.~\ref{diags}(E)--(G). However, due to problems with the complex tensor 
structure, the complete diagrams in Figs.~\ref{diags}(E)--(G) cannot be checked 
with this technique. 
 
In the next subsections, we explain the applications of these methods for our 
purposes and present a comparison between them.   
 
\subsection{Asymptotic expansions} 
 
We perform an expansion in a parameter $x$, where 
\begin{equation} 
x = \frac{M_Z^2}{m_t^2} \sim \frac{1}{4}.  
\end{equation}  
For any two-loop problem, there are four regions to consider.  
Let $k_1$ and $k_2$  
represent the internal momenta in the loops and $p$ stand for any external 
momentum, while $m$ generically denotes all masses 
that are small compared to $m_t$, $m<m_t$. In our case, $m = M_W,M_Z$. 
Then the four regions can be identified as follows:
 
\vspace{1ex} 
\begin{tabular}{ll} 
1) $k_1 \sim m_t $ and $k_2 \sim m_t  $ &  
(expansions in small parameters: $p$ and $m$)\\ 
2) $k_1 \sim m $ and $k_2 \sim m_t  $ & 
(expansions in small parameters: $p$, $k_1$ and $m$)\\ 
3) $k_1 \sim m_t $ and $k_2 \sim m  $ & 
(expansions in small parameters: $p$, $k_2$ and $m$) \\ 
4) $k_1 \sim m $ and $k_2 \sim m  $ & 
(expansions in small parameters: $p$, $k_1$, $k_2$ and $m$)  
\end{tabular} 
\vspace{1ex} 
 
This method allows us to represent two-loop vertex diagrams 
by a sum of simpler integrals, namely two-loop propagator and  vacuum integrals,  
plus one-loop integrals. However, higher orders in the expansion lead to  
higher powers of propagator denominators in these integrals.  
This is not a problem for one-loop or vacuum integrals, as analytic  
relations are well known; for relations and references, see, for example,  
Ref.~\cite{Smirnov:2004ym}. 
For two-loop propagator integrals, we employ  
the Laporta algorithm, as proposed in Ref.~\cite{laporta}.  
This algorithm allows us to automatically reduce  
complicated multi-loop integrals with non-trivial numerators  
to a smaller set of master integrals with unit numerators.  
In addition to the well-known integration by parts relations \cite{ibp}, 
Lorentz identities \cite{lor} can be used for faster performance. 
In our approach, the Laporta integral reduction is accomplished with the help
of the program {\sc IdSolver} \cite{Awramik:2004qv,idsolver}.   
 
As already observed for the two-loop vertex diagrams in the leptonic case 
\cite{sineffTot}, this expansion has a fast convergence behaviour. After 
performing the expansion down to fifth order, $\mathcal{O}(x^5)$, a precision of $10^{-5}$
in the final result for the two-loop part of $\sin^2\theta_{\mathrm{eff}}^{b\overline{b}}$ is obtained.  
 
\subsection{Semi-numerical integration based on the BT algorithm} 
 
Any Feynman diagram can be described by an integral 
\begin{equation} 
  \int_{\mathcal{S}}\d\x\,
\mathcal{Q}(\x) \prod_i \mathcal{V}_{i}^{\mu_i}(\x), 
\label{eq:FI} 
\end{equation}  
where $\mathcal{Q}(\x)$ and $\mathcal{V}_i(\x)$ are polynomials of {\bf x}, 
$\mathcal{Q}$ is the numerator of a Feynman integral, $\mathcal{V}_i$ is a denominator of  
the Feynman integral with the power 
$\mu_i$,  
which depends on $\varepsilon$, 
$\mathrm{\bf x}= (x_1,...,x_n)$ represents the  
$n$-dimensional vector space of Feynman parameters, and 
${\cal S}$ is the integration region defined by 
\begin{equation} 
\int_{\mathcal{S}}\d\x =  
\int_{0}^{1}\d x_1\int_{0}^{1-x_1}\d x_2\ldots  
\int_{0}^{1-\sum_{i=1}^{n-1}x_i}\d x_n. 
\end{equation} 
Tkachov proved the existence of an algorithm \cite{Tkachov},  later called 
BT algorithm, that can transform a Feynman integral of the type 
in Eq.~(\ref{eq:FI}) into a form  with better arrangement of divergences, which is 
consequently more suitable for numerical integration.  However, until now, 
compact-form solutions of this algorithm are only known at the one-loop order, which 
will be described in the following. 
 
For one-loop cases, $\mathcal{V}(\x)$ is a quadratic polynomial of {\bf x} of the form 
\begin{equation}  
  \mathcal{V}(\x) = \x^T \mathcal{W}  \x + 2 \R^T \x  + Z,  
\end{equation}   
where $\mathcal{W}$ is a $n\times n$ matrix, {\bf R} is a $n$-dimensional vector 
and $Z$ is a scalar number. Then one can show that the following relation  
is fulfilled:  
\begin{equation} 
\label{eq:BTrelation1l} 
\mathcal{V}^{\mu} = \frac{1}{\Delta} \left( 1 - \frac{(\x+\A)
{\mbox{\boldmath$\partial$}}}{2 (\mu + 1)}\right)  
\mathcal{V}^{\mu+1},  
\end{equation} 
where  
$\Delta = (Z - \R^T \mathcal{W}^{-1} \R)$ and  
$\A = \R^T \mathcal{W}^{-1}$.  
By application of this one-loop BT relation, supplemented by  
integration-by-parts identities, the power of the polynomial $\mathcal{V}(\x)$ of the 
Feynman integral is raised.   
For example, for the one-loop three-point function in three dimensions, one
finds
\begin{eqnarray} 
\lefteqn{ 
\int_0^1 \d x_1  
\int_0^{1-x_1} \d x_2  
\int_0^{1-x_1-x_2} \d x_3 
\, \mathcal{Q}(\x) \mathcal{V}^{\mu}(\x)}\nonumber\\ 
&=& 
\int_0^1 \d x_1  
\int_0^{1-x_1} \d x_2  
\int_0^{1-x_1-x_2} \d x_3 
\frac{1}{\Delta}\mathcal{V}^{\mu+1}(\x)\nonumber\\ 
&&{}\times  
\left\{
\mathcal{Q}(\x)+\frac{1}{2(\mu+1)}\sum_{k=1}^{3}\frac{\partial}{\partial x_k}
[(x_k+A_k) \mathcal{Q}(\x)] 
\right\} \nonumber\\ 
&&{}- 
\int_0^1 \d x_1  
\int_0^{1-x_1} \d x_2 
\frac{1}{\Delta}\,\frac{1}{2(\mu+1)} 
\left(1+\sum_{k=1}^3A_k\right)\mathcal{Q}(\x)\mathcal{V}^{\mu+1}(\x)|_{x_3=1-x_1-x_2}\nonumber\\ 
&&{}+  
\int_0^1 \d x_1  
\int_0^{1-x_1} \d x_2 
\frac{1}{\Delta}\,\frac{1}{2(\mu+1)}A_3\mathcal{Q}(\x)\mathcal{V}^{\mu+1}(\x)|_{x_3=0}\nonumber\\ 
&&{}+  
\int_0^1 \d x_1  
\int_0^{1-x_1} \d x_3 
\frac{1}{\Delta}\,\frac{1}{2(\mu+1)}A_2\mathcal{Q}(\x)\mathcal{V}^{\mu+1}(\x)|_{x_2=0}\nonumber\\ 
&&{}+  
\int_0^1 \d x_2  
\int_0^{1-x_2} \d x_3 
\frac{1}{\Delta}\,\frac{1}{2(\mu+1)}A_1\mathcal{Q}(\x)\mathcal{V}^{\mu+1}(\x)|_{x_1=0}. 
\end{eqnarray} 
This step can be applied iteratively until the power of $\mathcal{V}$ is as high as required,  
optimally $\mu = - n + \epsilon \to \epsilon$, where $n$ is a positive integer.  
In this way, the original integral is expressed in terms of a sum of different  
integrals, which possess better integration properties.   
 
Although no general and compact-form solution of the BT algorithm is 
known  for problems beyond the one-loop case,  it is only 
natural to apply a one-loop BT relation to a sub-loop of a two-loop integral.  
In this way, the integral can be made smooth with respect to the Feynman 
parameters  connected with the sub-loop to which the BT procedure is applied.
Due to the  
size of the expressions and their divergency structure, it is usually better to apply 
this relation to the sub-loop  with the highest number of internal lines.  
 
Finally, the $\epsilon$ poles in each component $x_i$ 
of the vector {\bf x}
are extracted by the relation  
\begin{eqnarray} 
\label{eq:subtraction} 
\lefteqn{\int_0^1 \d x_i\, x_i^{-n+\epsilon} f(\x)}\nonumber \\ 
&=&\int_0^1 \d x_i \,x_i^{-n+\epsilon}  
\left(f(\x)- \sum_{k=0}^{n-1}x_i^{k} \frac{f^{(k)}(0)}{k!} \right) 
+\sum_{k=0}^{n-1}  \frac{f^{(k)}(0)}{k! (k+1-n+\epsilon)}, 
\end{eqnarray} 
where $i=1,\ldots,n$.
 
\subsection{Semi-numerical integration based on dispersion relations} 
 
This method makes use of the fact that the one-loop self-energy can be written, 
with the help of a dispersion relation, as an integral over an expression that 
has the analytical form of a propagator  
\begin{eqnarray} 
B_0(p^2,m_1^2,m_2^2) &=& \int_{(m_1+m_2)^2}^\infty {\rm d}s 
  \frac{\Delta B_0(s,m_1^2,m_2^2)}{s - p^2},
\nonumber \\ 
\Delta B_0(s,m_1^2,m_2^2) &=& (4\pi\mu^2)^{4-D} 
  \frac{\Gamma(D/2-1)}{\Gamma(D-2)} \,
 \frac{\lambda^{(D-3)/2}(s,m_1^2,m_2^2)}{s^{D/2-1}},
\end{eqnarray}
where $\lambda(x,y,z)=x^2+y^2+z^2-2(xy+yz+zx)$.
If this one-loop self-energy is a sub-loop 
of a two-loop integral, the dispersion relation effectively reduces this integral 
to a one-loop integral with the additional integration over $s$, which  is performed 
numerically \cite{intnum}. 
 
Integrals with sub-loop triangles can be reduced to integrals with sub-loop 
self-energies by introducing Feynman parameters \cite{feynpar}. The integration 
over the Feynman parameters is also performed numerically. More details can be 
found in Ref.~\cite{sineffTot}. If the two-loop integrals contains ultraviolet,
infrared or 
threshold divergences, they need to be subtracted before the numerical 
integration can be carried out. For our purposes, this is achieved 
by subtracting a term from the integrand that can be integrated analytically. 
 
The reduction of integrals with 
irreducible numerators to a small set of master integrals is accomplished by 
using integration-by-parts and Lorentz-invariance identities. For complex 
diagrams with triangle sub-loops, the number of required relations  
can become very large, which is a limitation of this method. Therefore we do
not use it to compute the complete result for the two-loop corrections to 
$\sin^2\theta_{\mathrm{eff}}^{b\overline{b}}$, but only for cross checks of individual integrals and diagrams. 
%
 
\subsection{Comparison of methods} 
%
\begin{table}[tb] 
  \begin{center} 
    \begin{tabular}{lllll} 
      \hline 
      Diagram & Method & Result [$(\alpha/4 \pi)^2$]&&\\  
      \hline  
      (A) & semi-numerical    & $+3.82775120/\epsilon^2 $ & $ + 3.88128795/\epsilon $ & $ - 19.1983330 $ \\ 
          & $m_t$ expansion & $+3.82775120/\epsilon^2 $ & $ + 3.88/\epsilon      $ & $ - 19.19$ \\ 
      \hline  
      (B) & semi-numerical    & $+3.82775120/\epsilon^2 $ & $ -8.67823832/\epsilon  $ & $ + 25.4468576 $ \\ 
          & $m_t$ expansion & $+3.82775120/\epsilon^2 $ & $ -8.68/\epsilon       $ & $ + 25.45$ \\ 
      \hline  
      (C) & semi-numerical & $0/\epsilon^2     $ & $ + 0.90521614/\epsilon $ & $- 0.60580110$ \\ 
          & $m_t$ expansion & $0/\epsilon^2  $ & $ + 0.905/\epsilon      $ & $- 0.61$ \\ 
      \hline  
      (D) & semi-numerical    & $0/\epsilon^2 $ & $ + 1.55085212/\epsilon $ & $ - 4.50488822$ \\ 
          & $m_t$ expansion & $0/\epsilon^2 $ & $ + 1.55/\epsilon $ & $ - 4.50$ \\  
      \hline  
      (E) & semi-numerical &    $-2.30183413/\epsilon^2 $ & $ + 5.07108758/\epsilon $ & $+ 8.32594367$ \\ 
          & $m_t$ expansion & $-2.30183413/\epsilon^2 $ & $ + 5.07/\epsilon $ & $+ 8.33$ \\ 
      \hline  
      (F) & semi-numerical & $-2.80183413/\epsilon^2 $ & $ + 6.17261951/\epsilon $ & $ - 14.028 $ \\ 
      & $m_t$ expansion &  $-2.80183413/\epsilon^2 $ & $ + 6.17/\epsilon      $ & $ - 14.03 $ \\ 
      \hline  
      (G) & semi-numerical &    $-1.80183413/\epsilon^2    $ & $ + 3.9695556(5)/\epsilon $ & $ - 13.539 $ \\ 
          & $m_t$ expansion & $-1.80183413/\epsilon^2 $ & $ + 3.97/\epsilon $ & $ - 13.54 $ \\ 
      \hline 
    \end{tabular} 
  \end{center} 
  \vspace{1ex} 
  \caption{Comparison of the top-quark mass expansion with semi-numerical integrations for  
    selected diagrams of Fig.~\ref{diags}. For diagrams~(A)--(D),
    the semi-numerical results for 
    the BT and dispersion relation methods agree to all given digits, while for
    diagrams (E)--(G) results are available only for the BT method.}  
  \label{tab:compDiags} 
\end{table} 
%
In this section, we compare our three methods, based on  the top-quark mass expansion 
algorithm, the BT method, and the dispersion relations, where applicable. For the 
comparison, we use the following dimensionless input parameters:  $M_Z=1$, 
$M_W=80/91$, $m_t=180/91$, and the scale for dimensional regularisation 
$\mu = e^{\gamma_{\mathrm{E}}}$. The large-mass expansion is performed 
down to  ${\cal O}(m_t^{-12})$.   The expressions are normalised as they enter in 
$\sin^2\theta_{\mathrm{eff}}^{b\overline{b}}$, with the common prefactor $(\alpha/4 \pi)^2$ factored out.  For the 
comparison, we selected the set of diagrams corresponding to  Fig.~\ref{diags}, 
where only $W$-boson propagators in Feynman gauge, but not Goldstone-boson 
exchange has been included. For light-fermion loops, results are shown for 
leptons in the sub-loops, summed over the three lepton families, and for 
diagram (G) we chose the $ll\nu$ sub-loop, also summed over generations. 
Our results are presented in Table~\ref{tab:compDiags}. 
 
Where available, the results from the BT method and the method based on  
dispersion relations agree to all digit shown in the table.
As mentioned above, no complete results for diagrams (E)--(G) could be 
obtained with the dispersion relation method. 
 
There are clear advantages to the use of  
large-$m_t$ asymptotic expansions. No special treatment is required 
for the different types of divergences.  
Each large-mass pattern is associated with one expansion scheme,  
which is not sensitive to threshold divergences  
produced by the presence of small masses. In effect, the resulting programs 
are general and concise.  
As can be seen from Table~\ref{tab:compDiags},  
the obvious drawback of this method is the limited precision  
of the final results.  
However, it was observed that, with reasonable investment of computer 
time, the precision can be pushed to as high as required by the problem at hand.    

The semi-numerical programs usually produce results of better precision.  
However, the iterative application of the BT relation should be kept at a
minimum, as otherwise the precision  
can actually be lost. 
At least in our realization, the semi-numerical programs based on the BT and 
dispersion relation methods are  
not as general as the expansion technique. Special care has to be taken to  
deal with threshold-divergent cases, and the size of intermediate expressions 
can actually be much larger than what is observed during the use of the 
large-mass expansion.  
In the end, the time required for the analytic simplifications of these semi-numerical  
programs  
can be as large as the time required for performing a high-order large-mass expansion.   

It should also be kept in mind that the algorithms for  asymptotic expansions 
can be easily generalised to higher orders in perturbation theory, and were 
already applied for three-loop problems; see, for example, 
Refs.~\cite{threeloopexp,mt6}. The application of the
BT method in such cases  is more complicated, and no physical 
results are known at this moment.   
     
\section{Results} 
\label{results} 
 
%
\begin{table}[tb] 
\begin{center} 
\begin{tabular}{ll} 
\hline 
      Input parameter & Value\\  
      \hline  
      $M_W$ & $(80.404 \pm 0.0030)$~GeV \\ 
      $M_Z$ & $(91.1876 \pm 0.0021)$~GeV \\ 
      $\Gamma_Z$ & 2.4952~GeV \\ 
      $m_t$ & $(172.5 \pm 2.3)$~GeV \\ 
      $m_b$ & 4.85~GeV \\ 
      $\Delta\alpha(M_Z)$ & $0.05907 \pm 0.00036$ \\ 
      $\alpha_s(M_Z)$ & $0.119 \pm 0.002$ \\ 
      $G_\mu$ & $1.16637 \times 10^{-5}$~GeV$^{-2}$ \\ 
\hline 
\end{tabular} 
\end{center} 
\vspace{-1ex} 
\caption{Experimental input parameters used in the numerical evaluation, from 
Refs.~\cite{lepewwg,Eidelman:2004wy}. 
\label{tab:input}} 
\end{table} 
%
The computational methods described in the previous section were
implemented in automatised computer codes, to be able to handle the large-size 
expressions at intermediate stages. 
The analytic algorithm for the BT reduction was written in 
FORM \cite{Vermaseren:2000nd} and {\sc Mathematica}.  
The code for the large-top-quark-mass expansion was also implemented in FORM. 
When necessary, 
the reduction of two-loop propagators with higher powers of  
numerators and denominators, which inevitably appear for higher orders of the 
asymptotic expansion, was performed with the program {\sc IdSolver}
\cite{Awramik:2004qv,idsolver}. 
For the problem at hand, it had to create and solve a set of several thousands of equations, 
which it can achieve with very little computing time. 
For the numerical integrations, we developed a fast code written in C 
with the help of the {\sc Cuba} library \cite{Hahn:2004fe}. 
 
The results presented here were tested at different levels.  
We checked the finiteness and gauge invariance  
of the two-loop contributions to $\sin^2\theta_{\mathrm{eff}}^{b\overline{b}}$ analytically. 
The numerical results for the new diagrams were tested  
with two different methods, as presented in the previous section.  
In addition, full evaluation was performed independently by  
two groups within our collaboration.  
 
In the following, we show our numerical results for the input parameters
listed in 
Table~\ref{tab:input}. For the sake of easy comparison, we use the same 
parameters as in previous publications on  $\sin^2\theta_{\mathrm{eff}}^{\mathrm{lept}}$ \cite{sineffTot},  which is 
justified by the fact that the changes of measured values are insignificant for 
this presentation.   It is important to note that the experimental values for 
the $W$- and $Z$-boson masses correspond to a Breit-Wigner parametrisation with a 
running width and have to be translated to the pole-mass scheme used in the 
loop calculations~\cite{muon}. In effect, this translation results in a downward 
shift \cite{riemann} of $M_Z$ and $M_W$ by about 34 and 28~MeV, 
respectively.
The non-zero mass of the bottom quark was retained in the 
${\cal{O}}(\alpha)$ contribution, but neglected in the two-loop part.  
%
\begin{figure}[htb] 
  \begin{center} 
    \psfig{figure=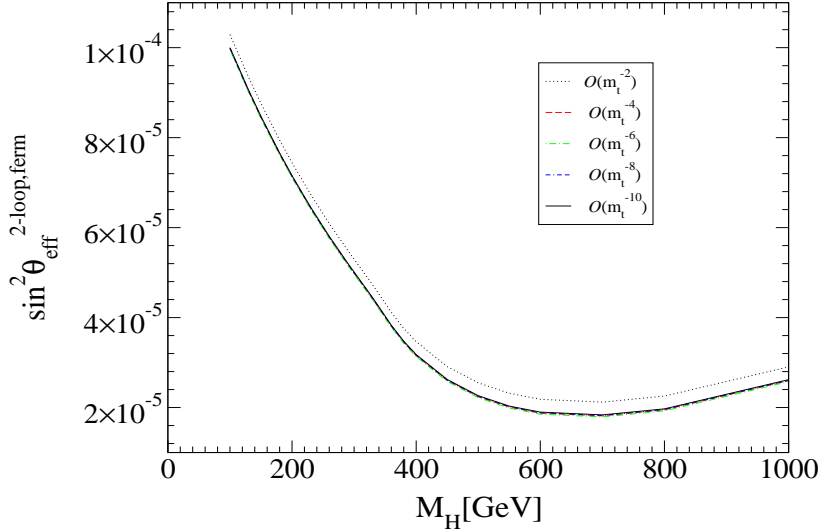, width=8cm,height=11cm,angle=270}  
  \end{center} 
  \caption{ 
    Two-loop fermionic contribution to $\sin^2\theta_{\mathrm{eff}}^{b\overline{b}}$, with the $W$-boson mass evaluated from the Standard Model and the
other input parameters taken from Table~\ref{tab:input}. 
    \label{fig:sinExpanded}} 
\end{figure} 
%
The complete fermionic two-loop contribution to $\sin^2\theta_{\mathrm{eff}}^{b\overline{b}}$ is shown in  
Fig.~\ref{fig:sinExpanded}, for various levels of the large-top-quark-mass
expansion.  For 
the plot, $G_\mu$ is used as an input parameter, and $M_W$ is calculated from it according to 
Eq.~(\ref{eq:delr}). As can be seen from the figure, the asymptotic expansion is 
converging robustly over the entire range of relevant Higgs-boson mass values. The 
relative error estimated for the expansion up to ${\cal{O}}(m_t^{-10})$ is 
$10^{-5}$ and thus more than sufficient for our purposes. The numerical values for 
selected values of the  Higgs-boson mass are shown in Table~\ref{tab:sinValues}. 
The left table uses $G_\mu$ as an input via Eq.~(\ref{eq:delr}), while a fixed 
mass $M_W=80.404$~GeV is used for the right table. For small Higgs-boson 
mass, the new correction is relatively large,  about $10^{-4}$. For larger
values of $M_H$, it 
can be up to one  order of magnitude smaller, about $10^{-5}$. 
%
\begin{table}[tb] 
  \centering
    \begin{tabular}{lll} 
    \hline 
    $M_H$ & ${\cal{O}}(\alpha)$ &  ${\cal{O}}(\alpha^2)_{\mathrm{ferm}}$\\[-1ex]  
    [GeV] & $[10^{-4}]$& $[10^{-4}]$ \\ 
    \hline  
    100  &  104.77  &  1.00 \\ 
    200  &  100.15  &  0.71 \\ 
    400  &  94.397  &  0.32 \\ 
    600  &  90.666  &  0.19 \\ 
    1000 &  85.748  &  0.26 \\ 
    \hline 
  \end{tabular} 
  \hspace{2em}
  \begin{tabular}{lll} 
    \hline 
    $M_H$ & ${\cal{O}}(\alpha)$ &  ${\cal{O}}(\alpha^2)_{\mathrm{ferm}}$\\[-1ex]  
    $[\mathrm{GeV}]$ & $[10^{-4}]$ & $[10^{-4}]$ \\ 
    \hline  
    100  &  105.03   &  0.98 \\ 
    200  &  100.74   &  0.67 \\ 
    400  &   95.354  &  0.24 \\ 
    600  &   91.847  &  0.10 \\ 
    1000 &   87.196  &  0.16 \\ 
    \hline 
  \end{tabular} 
  \caption{Contributions to $\sin^2\theta_{\mathrm{eff}}^{b\overline{b}}$,  
    with the $W$-boson mass evaluated from the Standard Model (left) 
    or fixed (right).} 
  \label{tab:sinValues} 
\end{table} 
%
Following earlier publications on two-loop electroweak corrections, 
we express our results in terms of fitting formulas.  
The form factor $\Delta\kappa^{(\alpha^2,\mathrm{ferm})}_{b\overline{b}}$,
which contains the fermionic two-loop electroweak corrections to
$\sin^2\theta_{\mathrm{eff}}^{b\overline{b}}$ according to  
Eq.~\eqref{eq:sin}, can be approximated as 
\begin{eqnarray} 
\label{eq:formkap} 
\Delta\kappa^{(\alpha^2,\mathrm{ferm})}_{b\overline{b}} &=&
 \Delta\alpha \Delta\kappa^{(\alpha)}_{b\overline{b}} + 
\Delta\kappa^{(\alpha^2,\mathrm{ferm})}_{b\overline{b},\mathrm{rem}},
\nonumber\\
\Delta\kappa^{(\alpha^2,\mathrm{ferm})}_{b\overline{b},\mathrm{rem}} &=&  
k_0 + k_1 L_H + k_2  L_H^2 + k_3  L_H^4 + k_4  (\Delta_H^2 -1)  
 + k_5  \Delta_t 
\nonumber\\
&&{} + k_6  \Delta_t^2  
 + k_7  \Delta_t L_H + k_8 \Delta_W + k_9 \Delta_W \Delta_t 
 + k_{10} \Delta_Z, 
\end{eqnarray} 
where $\Delta\kappa^{(\alpha)}_{b\overline{b}}$ is the one-loop result, and 
\begin{align} 
L_H &= \ln\frac{M_H}{100~\mathrm{GeV}}, & 
\Delta_H &= \frac{M_H}{100~\mathrm{GeV}}, & 
\Delta_t &= \left(\frac{m_t}{178~\mathrm{GeV}}\right)^2 -1,  
 \nonumber \\[1ex] 
\Delta_Z &= \frac{M_Z}{91.1876~\mathrm{GeV}} -1, &
\Delta_W &= \frac{M_W}{80.404~\mathrm{GeV}} -1. 
\end{align} 
Fitting this formula to the exact result, we obtain 
\begin{equation} 
\begin{aligned} 
k_0 &= -2.666\times 10^{-3}, & k_1 &= -5.92 \times 10^{-5}, & k_2 &= -3.29\times 10^{-6},  
\\ 
k_3 &= 3.49\times 10^{-6}, & 
k_4 &= 2.83\times 10^{-6}, & k_5 &= -5.34\times 10^{-3}, \\ 
k_6 &= -2.10\times 10^{-3}, & k_7 &= -2.19\times 10^{-4}, & 
k_8 &= -6.31\times 10^{-2}, \\ 
k_9 &= -1.26\times 10^{-1}, & k_{10} &= 6.47\times 10^{-2}. 
\end{aligned} 
\end{equation} 
This parametrisation reproduces  the exact calculation with maximal and average 
deviations  of $1.4 \times 10^{-5}$ and $5 \times 10^{-6}$, respectively, as  
long as the input parameters stay within their $2 \sigma$ ranges of the 
experimental errors quoted in 
Table~\ref{tab:input} and the Higgs-boson mass is in the range 10~GeV $\leq M_H 
\leq$ 1~TeV. If the top-quark mass and the $W$-boson mass vary within 4$\sigma$ 
ranges, the formula is still accurate to $2.1 \times 10^{-5}$. 
 
We also present a simple parametrisation for the currently best prediction for  
$\sin^2\theta_{\mathrm{eff}}^{b\overline{b}}$, including all known corrections to 
$\Delta\kappa_{b\overline{b}}$ and $\Delta r$ 
(for the calculation of $M_W$ from $G_\mu$ see Refs.~\cite{Freitas:2000gg,mw}). 
For $\Delta\kappa_{b\overline{b}}$, in addition to the one-loop and fermionic
two-loop electroweak corrections, we include QCD corrections of
${\cal{O}}(\alpha\alpha_s)$ \cite{qcd2} and ${\cal{O}}(\alpha\alpha_s^2)$
\cite{threeloopexp,qcd3} to the one-loop contribution, as well as universal
corrections for large top-quark mass, of ${\cal{O}}(\alpha^2\alpha_s m_t^4)$
and ${\cal{O}}(\alpha^3m_t^6)$ \cite{mt6}. 
Moreover, leading four-loop QCD correction to the $\rho$ parameter, which 
arise from top- and bottom-quark loops, are taken into account
\cite{Chetyrkin:2006bj}.
We use the parametrisation 
\begin{eqnarray} 
\label{eq:formula} 
\sin^2\theta_{\mathrm{eff}}^{b\overline{b}}& =& s_0 + d_1 L_H + d_2  L_H^2 + d_3  L_H^4 + d_4  (\Delta_H^2 -1)  
 + d_5  \Delta_\alpha
\nonumber\\
&&{}+ d_6  \Delta_t + d_7  \Delta_t^2  
 + d_8  \Delta_t  (\Delta_H -1) 
 + d_9  \Delta_{\alpha_s} + d_{10} \Delta_Z, 
\end{eqnarray} 
with 
\begin{equation} 
\Delta_\alpha = \frac{\Delta \alpha(M_Z)}{0.05907}-1, \qquad 
\Delta_{\alpha_s} = \frac{\alpha_s(M_Z)}{0.117}-1. 
\end{equation} 
The best-fit numerical values for the coefficients are 
\begin{equation}
\begin{aligned}
s_0 &= 2.327580\times10^{-1}, &
d_1 &= 4.749\times 10^{-4}, &
d_2 &= 2.03\times 10^{-5}, \\
d_3 &= 3.94\times 10^{-6}, &
d_4 &= -1.84\times 10^{-6}, &
d_5 &= 2.08\times^{10^-2}, \\
d_6 &= -9.93\times 10^{-4}, &
d_7 &= 7.08\times 10^{-5}, &
d_8 &= -7.61\times 10^{-6}, \\
d_9 &= 4.03\times 10^{-4}, &
d_{10} &= 6.61\times 10^{-1}.
\end{aligned}
\end{equation}
This parametrisation approximates the full result with  
maximal and average 
deviations  of $4.3 \times 10^{-6}$ and $1.3 \times 10^{-6}$, respectively, 
for 10~GeV${}\leq M_H  
\leq 1$~TeV and the other input parameters in their 
$2 \sigma$ ranges. 
 
\section{Conclusions} 
\label{concl} 
 
In this paper, the calculation of the two-loop electroweak fermionic corrections 
to the effective weak-mixing angle for the $Zb\bar{b}$ vertex, $\sin^2\theta_{\mathrm{eff}}^{b\overline{b}}$, was
presented. Such an accurate theoretical prediction for $\sin^2\theta_{\mathrm{eff}}^{b\overline{b}}$ is necessary 
for the interpretation of the bottom-quark asymmetry measurements at the 
$Z$-boson pole. Compared to the previously known corrections to $\sin^2\theta_{\mathrm{eff}}^{b\overline{b}}$, 
the new electroweak two-loop result turns out to be sizable, of order ${\cal 
O}(10^{-4})$ for a Higgs-boson mass near 100~GeV. 
 
The calculation was performed by using methods that had been used earlier for 
the computation of the leptonic effective weak-mixing angle, as well as a 
newly developed code based on the BT algorithm. The results of 
the different methods were checked against each other. 
 
Although we did not perform a detailed analysis of 
the error from unknown high-order corrections, in particular the missing bosonic 
two-loop corrections and terms of order ${\cal 
O}(\alpha^2\alpha_s)$, we expect those to be of similar order as for the leptonic 
effective weak-mixing angle. The main difference between the leptonic and 
bottom-quark effective weak-mixing angles are the vertex diagrams with internal 
$W$-boson and top-quark propagators. While leading to numerical differences between 
$\sin^2\theta_{\mathrm{eff}}^{\mathrm{lept}}$ and $\sin^2\theta_{\mathrm{eff}}^{b\overline{b}}$, 
these diagrams do not introduce special enhancement or 
suppression factors. Therefore, we expect the theoretical uncertainty to our 
result for $\sin^2\theta_{\mathrm{eff}}^{b\overline{b}}$ to be about
$5 \times 10^{-5}$, similar to Ref.~\cite{sineffTot}.

\section*{Acknowledgements} 
 
The work of M.A. and B.A.K. was supported in part by the German Research 
Foundation (DFG) through Grant No.\ KN~365/3-1 and through the Collaborative
Research Centre No.~676
{\it Particles, Strings and the Early Universe --- the structure of Matter and
Space Time}.
The work of M.C. was supported in part by the Sofja Kovalevskaja Award of the  
Alexander von Humboldt Foundation  
and by the ToK Program {\it ALGOTOOLS} (MTKD-CD-2004-014319). 
A.F. is grateful for warm hospitality at Argonne National Laboratory and the
Enrico Fermi Institute of the University of Chicago, where part of his work on
this project was performed. 
 
 

\end{document}